\documentclass[11pt,onecolumn,amssymb,nofootinbib]{revtex4}
\usepackage{amsmath, amsthm, amscd, amssymb}
\usepackage{bm}
\usepackage{bbm}

\begin{document}

\title{\bf Coleman-Weinberg symmetry breaking in $SU(8)$ induced by
a third rank antisymmetric tensor scalar field II: the fermion spectrum}

\author{Stephen L. Adler}
\email{adler@ias.edu} \affiliation{Institute for Advanced Study,
Einstein Drive, Princeton, NJ 08540, USA.}

\begin{abstract}
We continue our study of Coleman-Weinberg symmetry breaking induced by a third rank antisymmetric tensor scalar, in the context of the  $SU(8)$ model \cite{adler1} we proposed earlier.  We focus in this paper on qualitative features that will determine whether the model can make
contact with the observed particle spectrum.  We discuss
the mechanism for giving the spin $\frac{3}{2}$ field a mass by the  BEH mechanism, and analyze the remaining massless spin $\frac{1}{2}$
fermions, the global chiral symmetries, and the running couplings after symmetry breaking.  We note that the smallest gluon
mass matrix eigenvalue has an eigenvector suggestive of $U(1)_{B-L}$, and conjecture that the theory runs to an infrared
fixed point at which there is a massless gluon with 3 to -1 ratios in generator components.   Assuming this, we discuss a mechanism for making
contact with the standard model,  based on a conjectured asymmetric breaking of $Sp(4)$ to $SU(2)$ subgroups, one of which is the electroweak
$SU(2)$, and the other of which is a ``technicolor'' group that binds the  original $SU(8)$ model fermions, which  play the role of ``preons'',
into composites.
Quarks can emerge as 5 preon composites and leptons as 3 preon composites, with consequent stability of the proton against decay to a single lepton plus a meson.  A composite Higgs boson can emerge as a two preon composite.  Since  anomaly matching  for the
relevant conserved global symmetry current is not obeyed by three fermion families, emergence of three composite families requires formation of a Goldstone boson with quantum numbers matching this current, which can be a light dark matter candidate.
\end{abstract}

\maketitle

\section{Introduction}

In this paper we continue our study of Coleman-Weinberg symmetry breaking in $SU(8)$ induced by a third rank antisymmetric
tensor scalar field, which was initiated in an earlier paper \cite{adler3}.  In that paper, referred to hereafter as (I), we showed that
$SU(8)$ is broken to $SU(3)\times Sp(4)$, and we gave a detailed analysis of the Goldstone boson structure and the BEH mechanism, together with a group-theoretic classification of residual states after symmetry breaking. The explicit numerical calculation of (I) shows that the 56 of scalars responsible for $SU(8)$ symmetry
breaking leads to no residual scalar states below the symmetry breaking scale, since all components of the 56 are either absorbed by the
BEH mechanism into longitudinal components of the vector mesons, or  obtain masses at the symmetry breaking scale.  However, the questions of the fermion
spectrum after symmetry breaking, and the absence of a massless $U(1)$ gauge boson, were not addressed.  We begin the analysis
of these issues in this paper.  In Sec. 2 we discuss the BEH mechanism for giving a mass to the spin $\frac{3}{2}$ field, and in Sec. 3
we analyze the global chiral symmetries of the model, taking $SU(8)$  instantons into account.   In
Sec. 4 we analyze the residual massless fermions, and in  Sec. 5
the $SU(3)$ and $Sp(4)$ running couplings, after $SU(8)$ symmetry breaking.  In Sec. 6 we begin the study of how the $SU(8)$ model of \cite{adler1} can make contact with the fermion spectrum of the standard model.  In Sec. 6A
we discuss mechanisms for generating the weak hypercharge, and their possible relation to the structure of the smallest
mass matrix eigenvalue found in (I).  In Sec. 6B we state three conjectures about the behavior of the $SU(8)$ theory with full
radiative corrections, and based on these conjectures discuss constraints coming from extrapolation of the standard model couplings in Sec. 6C, the $U(1)$ generator structure
of the 56 and $\overline{28}$ fermionic preons in Sec. 6D, and the counting of three preon and five preon candidates for composite
leptons and quarks in Sec. 6E. Enumeration of two preon candidates for a composite Higgs boson is given in Sec. 6F, and a possible
mechanism for giving a double-singlet three family structure is given in Sec. 6G.  In Sec. 7 we
analyze 't  Hooft  anomaly matching conditions for our model.  In Sec. 8 we review motivations for and encouraging features of  the $SU(8)$ model, discuss further steps to be undertaken, and state some experimental consequences that follow from the qualitative analysis of this paper.

\section{BEH mechanism for the spin-3/2 field}
We turn now to consequences for the fermion spectrum arising from the scalar field
minimum $\overline{\phi}$ produced by Coleman-Weinberg symmetry breaking.     A key feature
of the model of  \cite{adler1} is that invariance under the $SU(8)$ group forbids Yukawa
couplings of the $56$ and $\overline{28}$ representation  spin $\frac{1}{2}$ fermions to the $56$ representation scalar field $\phi$.
However, as also pointed out in \cite{adler1}, $SU(8)$ allows a coupling of the representation $8$  spin $\frac{3}{2}$ fermion field to a linear
combination $\lambda$
of the $\overline{28}$ spin $\frac{1}{2}$ fermion fields $\lambda_a$, a=1,2 of the form
\begin{equation}\label{RS}
\overline{\lambda}^{[\alpha\beta]} \gamma^{\nu} \psi_{\nu}^{\gamma} \phi^*_{[\alpha \beta \gamma]}
\end{equation}
and its conjugate, which vanish when the gravitino gauge fixing condition
\begin{equation}\label{fix}
\gamma^{\nu} \psi_{\nu}^{\gamma} =0
\end{equation}
 is imposed.
In \cite{adler1} this vanishing was adduced as a reason for excluding the coupling term of Eq. \eqref{RS}, but in fact this is not correct
when the vector field gauge coupling $g$ is nonzero.  The coupling term of Eq. \eqref{RS} vanishes only in the ungauged $g=0$ limit, both
because then $ \gamma^{\nu} \psi_{\nu}^{\gamma} =0$ is a valid gauge condition, and also because in the zero gauge coupling limit,
there is  a supersymmetry \cite{adler1} linking the scalar $\phi$ and the fermions $\lambda_a$ under which the coupling of Eq. \eqref{RS}
is not invariant.

To see what happens when one attempts to impose the condition $ \gamma^{\nu} \psi_{\nu}^{\gamma} =0$ in the presence
of gauge couplings, we follow the analysis of \cite{adler2} and replace the four-component left chiral field $\psi_{\mu}^{\gamma}$ by its
two-component equivalent $\Psi_{\mu}^{\gamma}$, in terms of which the condition of Eq. \eqref{fix} takes the form
\begin{equation}\label{fix1}
\Psi_0^{\gamma}=\vec \sigma \cdot \vec \Psi^{\gamma}~~~,
\end{equation}
with $\vec \sigma$ the Pauli matrices.
Multiplying Eq. \eqref{fix1} from the left by $\vec \sigma \cdot \vec B$ (with $\vec B$, $\vec E$ the magnetic, electric gauge fields,
and from here on suppressing internal symmetry indices, which
are always understood to be contracted in the natural order, e.g.  $(\vec B \Psi_{\mu})^{\beta} \equiv  \vec B_{\gamma}^{\beta} \Psi_{\mu}^{\gamma}$ etc.)
we get after some Pauli matrix algebra,
\begin{equation}\label{fix2}
\vec \sigma \cdot \vec B \Psi_0=\vec \sigma \cdot \vec B  \vec \sigma \cdot \vec \Psi
=\vec B \cdot \vec \Psi+ i \vec \sigma \cdot \vec B \times \vec \Psi~~~.
\end{equation}
Comparing this with the secondary Rarita-Schwinger constraint, which in two-component form reads
\begin{equation}\label{sec}
\vec \sigma \cdot \vec B \Psi_0=\vec B \cdot \vec \Psi + \vec \sigma \cdot \vec E \times \vec \Psi~~~,
\end{equation}
we get
\begin{equation}\label{fix3}
\vec \sigma \cdot (\vec B + i\vec E)\times \vec \Psi~~~=0.
\end{equation}
In general this will conflict with the Rarita-Schwinger primary constraint
\begin{equation}\label{primary}
\vec \sigma \cdot \vec D \times \vec \Psi=0~~~,
\end{equation}
where $\vec D=\vec \nabla + g \vec A$ is the gauge field covariant derivative.  So in the case of
nonzero gauge coupling, Eq. \eqref{fix} is not a satisfactory gauge constraint on the spin $\frac{3}{2}$
field.  In \cite{adler2}, it is shown that the natural constraint for quantizing the Rarita-Schwinger
field, in the presence of gauge couplings, is in fact the gauge covariant radiation gauge condition
$\vec D \cdot \vec \Psi=0$.

In view of this analysis, a coupling term of the form of Eq. \eqref{RS} plus its adjoint  is allowed with a coefficient
that is a nonzero power of the gauge coupling $g$.  When the scalar field $\phi$ develops an expectation $\overline{\phi}$,
this leads to a self-adjoint coupling term proportional to
\begin{equation}\label{RS1}
\overline{\lambda}^{[\alpha\beta]} \gamma^{\nu} \psi_{\nu}^{\gamma} \overline{\phi}^{\,*}_{[\alpha \beta \gamma]}
-\overline{\psi}_{\nu\gamma}\gamma^{\nu}\lambda_{[\alpha\beta]}\overline{\phi}^{[\alpha\beta\gamma]}~~~.
\end{equation}
Writing
\begin{align}\label{phihat}
| \overline{\phi}|\equiv&\big[\overline{\phi}^{\,*}_{[\alpha \beta \gamma]}\overline{\phi}^{[\alpha\beta\gamma]}]^{1/2}~~~,\cr
\hat{{\phi}}^{[\alpha\beta\gamma]} = &\overline{\phi}^{[\alpha\beta\gamma]}/| \overline{\phi}|~~~,\cr
\end{align}
we can diagonalize Eq. \eqref{RS1}  into terms  proportional to
\begin{equation}\label{RS2}
 |\overline{\phi}|  [(\overline{\lambda}^{[\alpha\beta]} \hat{\phi}^{\,*}_{[\alpha\beta\gamma]}-\overline{\psi}_{\nu\gamma}\gamma^{\nu})
(\lambda_{[\alpha\beta]}\hat{\phi}^{[\alpha\beta\gamma]}+\gamma^{\nu}{\psi}^{\gamma}_{\nu})
-(\overline{\lambda}^{[\alpha\beta]} \hat{\phi}^{\,*}_{[\alpha\beta\gamma]}+\overline{\psi}_{\nu\gamma}\gamma^{\nu})
(\lambda_{[\alpha\beta]}\hat{\phi}^{[\alpha\beta\gamma]}-\gamma^{\nu}{\psi}^{\gamma}_{\nu})]~~~,
\end{equation}
with the dimension one coefficient $ |\overline{\phi}|$ giving a mass through the BEH mechanism which has a magnitude near the $SU(8)$ breaking scale.  Note that unlike a conventional Dirac mass term, which couples
left to right chiral fields, Eq. \eqref{RS2} couples left chiral fields to left chiral fields.  Nonetheless,  we expect it to alter the propagators for the spin $\frac{3}{2}$ field, and the components of the $\overline{28}$ spin $\frac{1}{2}$ fields
that enter into Eqs. \eqref{RS}, \eqref{RS1}, and \eqref{RS2}, in a way that removes them from the low energy spectrum.

\section{Global chiral symmetries}

We turn next to an analysis of the global chiral symmetries of the model of Ref. 1.  Taking into account the Lagrangian term of Eq. \eqref{RS},
the model is formally invariant under independent $U(1)$ rephasings of the representation 8 Rarita-Schwinger field $\psi_{\mu}$, the
representation 56 fermion field $\chi$, and the representation $\overline{28}$ fermion field $\lambda_{\perp}$ that is orthogonal to
the linear combination $\lambda$ of $\overline{28}$ fields that is coupled to the Rarita-Schwinger field through Eq. \eqref{RS}.  Rephasing
invariance of Eq. \eqref{RS} requires that $\lambda$ be assigned the same phase as $\psi_{\mu}$, and so when Eq. \eqref{RS} is included in the action the formal chiral symmetry group, before taking $SU(8)$ anomaly effects into account, is $U(1)^3$.

Each independent rephasing corresponds to a formally conserved current, calculated from the kinetic terms in the action,  with an associated conserved charge operator.   However, when all rephasings are taken as the same, the corresponding
current is the overall $U(1)$ current, which has an $SU(8)$ anomaly through the triangle diagram when quantum effects are included, and so is non-conserved .  Taking quantum effects into account, the effective global chiral symmetry group is reduced to $U(1)^2$.  To find the associated
conserved charges, we look for the anomaly-free linear combination $\hat Q$ of charges $N_8+N_{\overline{28}\lambda}$, $N_{56}$, and $N_{\overline{28}\perp}$ associated with the fields $\psi_{\mu}$, $\lambda$, $\chi$, and $\lambda_{\perp}$ respectively,
\begin{equation} \label{anomfree1}
\hat Q=K_{8} (N_{8}+N_{\overline{28}\lambda}) + K_{56} N_{56} + K_{\overline{28}\perp} N_{\overline{28}\perp}~~~.
\end{equation}
The anomaly associated with each $N$ in Eq. \eqref{anomfree1} is proportional to the index of the corresponding representation, times
an extra factor of 3 for Rarita-Schwinger fields.  Since the indices of the $SU(8)$ representations 8, 56, and $\overline{28}$ are
1, 15, and 6 respectively (see Table I and footnote 1 below), the condition for $\hat Q$ to be free of $SU(8)$ anomalies is
\begin{equation}\label{anomfree2}
0=9 K_{8} + 15 K_{56} + 6 K_{\overline{28}\perp}~~~,
\end{equation}
or dividing by a factor of 3,
 \begin{equation}\label{anomfree3}
0=3K_{8} + 5 K_{56} + 2 K_{\overline{28}\perp}~~~.
\end{equation}
Equation \eqref{anomfree3} has the following two independent solutions,
\begin{align}\label{solution}
K_8=&1,~K_{56}=-\frac{3}{5},~K_{\overline{28}\perp}=0~~~,\cr
K_8=&0,~K_{56}=1,~K_{\overline{28}\perp}=-\frac{5}{2}~~~,\cr
\end{align}
corresponding to the conserved $U(1)$ charges
\begin{align}\label{charges}
Q_1=&N_{8}+N_{\overline{28}\lambda}-\frac{3}{5} N_{56}~~~,\cr
Q_2=&N_{56}-\frac{5}{2} N_{\overline{28}\perp}~~~.\cr
\end{align}

Any linear combination of $N_{8}$, $N_{56}$, and $ N_{\overline{28}\perp}$  that cannot be written as a linear combination of $Q_1$ and $Q_2$ is non-conserved through
the $SU(8)$ anomaly, and gives rise to fermion non-conservation through instanton processes.  The basic instanton  process
can be represented as an effective Lagrangian coupled to a number of  field lines for each fermion equal to the field representation index \cite{shifman}, with an extra factor of 3 for spin-$\frac{3}{2}$ Rarita-Schwinger fermions.  Thus, an instanton process couples to 3 lines of the representation 8 field $\psi_{\mu}^{\alpha}$, 15 lines of the representation 56 field $\chi^{[\alpha\beta\gamma]}$,  6 lines of the representation $\overline{28}$
field $\lambda_{[\alpha\beta]}$, and 6 lines of the representation $\overline{28}$ field $\lambda_{\perp[\alpha\beta]}$, where we have now
shown the internal symmetry indices $\alpha,\,\beta,\,\gamma$.  As a check on the counting, the overall instanton process has $3+3\times 15 = 48$
upper indices, and $6 \times 2 + 6 \times 2=24$ lower indices, giving after contraction a net of 24 upper indices.  Since $24\equiv 0$ modulo 8,
these can be contracted to an internal symmetry singlet using three factors of the $SU(8)$ invariant totally antisymmetric tensor $\epsilon_{\alpha_1...\alpha_8}$.    Because the instanton
process involves three Rarita-Schwinger fields, which become heavy after $SU(8)$ breaking, and six fields $\lambda$, parts of which become
heavy as well, it cannot lead to fermion non-conserving effects
in purely low energy processes.  Thus, for low energy processes, in addition to $Q_2$ being conserved, the part of $Q_1$ involving
fields that do not become massive,
\begin{equation}\label{charges1}
\tilde{Q}_1=N_{\overline{28}\Delta\lambda}-\frac{3}{5} N_{56}~~~,
\end{equation}
is effectively conserved.  Here $\Delta \lambda$ denotes the part of $\lambda$ that does not mix with the Rarita-Schwinger fields through
the BEH mechanism, as discussed in further detail in the next section.
\section{Massless fermions remaining after $SU(8)$ breaking}

To see what massless fermions remain after $SU(8)$ breaking, we must analyze the coupling term of Eq. \eqref{RS} in terms
of the representations under the residual symmetry group $SU(3) \times Sp(4)$.  Recalling that
\begin{align}\label{decomp1}
8=&(1,4)+(3,1)+(1,1)~~~,\cr
\overline{28}=&(\overline{3},4)+(1,5)+(1,4)+(3,1)+(\overline{3},1)+(1,1)~~~,\cr
\end{align}
we see that the $(1,4)+(3,1)+(1,1)$ components of the spin $\frac{3}{2}$ field, which accounts for all of the
$SU(3) \times Sp(4)$ representation content of this field, couple to the corresponding components of the linear
combination of  $\overline{28}$ fields that enters into Eq. \eqref{RS}.  This leaves the following residual
massless fermion fields:

\begin{itemize}
\item  From the remainder $\Delta \lambda$ of the linear combination $\lambda$ of the two  $\overline{28}$ fermions
we have the representations
\begin{equation}\label{decomp2}
\Delta \lambda =(\overline{3},4)+(1,5)+(\overline{3},1)~~~.
\end{equation}
\item From the orthogonal linear combination $\lambda_{\perp}$ of the two $\overline{28}$ fermions we have the
representations
\begin{equation}\label{decompp1}
\lambda_{\perp}=(\overline{3},4)+(1,5)+(1,4)+(3,1)+(\overline{3},1)+(1,1)~~~.
\end{equation}
\item From the 56  fermions we have the representations
\begin{equation}\label{decompp2}
\chi=(3,5)+(3,4)+(\overline{3},4)+(3,1)+(\overline{3},1)+(1,5)+(1,4)+(1,1)_1+(1,1)_2~~~.
\end{equation}
\end{itemize}

Using Eqs. \eqref{decomp1}--\eqref{decompp2} we can verify that the $U(1)$ charges $Q_1$ and $Q_2$ of Eq. \eqref{charges}, which were constructed to be free of $SU(8)$ anomalies,
are also anomaly-free with respect to the $SU(8)$ subgroups $SU(3)$ and $Sp(4)$.  This is a consequence of a branching
sum rule for Lie algebra indices \cite{okubo} which can be stated as follows.  Let $L_1$ and $L_2$ be simple Lie algebras that are subgroups
of the algebra $L$, so that $L\supset L_1 \times L_2$.  A representation $R$ of $L$ will have the branching expansion
\begin{equation}\label{branch}
R= \sum_i (R_{1i},R_{2i})~~~,
\end{equation}
where we have used the notation of Eqs. \eqref{decomp1}--\eqref{decompp2}.
Then the index $c(R)$ of the representation $R$ is related to the indices $c(R_{1,2i})$ and dimensions $D(R_{1,2i})$ of the
representations $R_{1i}$, $R_{2i}$ by the sum rules,
\begin{equation}\label{branch1}
c(R)=\sum_i c(R_{1i})D(R_{2i}) = \sum_i D(R_{1i}) c(R_{2i})~~~.
\end{equation}
Taking $L_1$ as the Lie algebra $SU(3)$ and $L_2$ as the Lie algebra $Sp(4)$, and remembering to include an extra  factor of 3 multiplying Rarita-Schwinger indices, the first equality in Eq. \eqref{branch1} implies that
$Q_{1,2}$ are $SU(3)$ anomaly-free, and the second equality in Eq. \eqref{branch1} implies that $Q_{1,2}$ are $Sp(4)$ anomaly-free.
Analogous statements hold when later on we further decompose $Sp(4)$ into subgroups according to $Sp(4)\supset SU(2) \times SU(2)$.

\section{Running coupling analysis}

We turn next to a study of the non-Abelian running couplings for the various groups figuring in the model.  For any non-Abelian gauge theory, the running
coupling is governed by the equation $\mu dg/d\mu = \beta(g)$, where $\mu$ is the scale mass and the beta function $\beta(g)$ is given by
\begin{align}\label{beta}
\beta(g)=&-\frac{g^3}{32\pi^2}\frac{1}{3}K~~~, \cr
K \equiv&11 c(1)-26 c({\rm Weyl}\, 3/2)-2 c(\rm {Weyl}\, 1/2)-c({\rm complex}\, 0)~~~,\cr
\end{align}
with $c(s)$ the index of the group representation with spin $s$ in the conventions\footnote{The customary particle physics normalization
uses index values that are one-half of those in the McKay-Patera tables, and thus has the $32\pi^2$  in the denominator of Eq. \eqref{beta} replaced by $16 \pi^2$.  I am indebted to Peter
Goddard for explaining the connection \cite{goddard} between index normalization and the normalization convention for the longest group weight, and to Paul Langacker for pointing
out that the Slanksy review \cite{slansky}  uses the McKay-Patera index values without making the corresponding change in the denominator of Eq. \eqref{beta}.}  of the McKay-Patera tables  \cite {mckay}.  As noted in \cite{adler1}, before $SU(8)$ breaking the theory has $K=81$, obtained by substituting the index values from Table I into Eq. \eqref{beta}, giving
\begin{equation}
K=11(16)-26(1) -2(15 + 2 \times 6) -15= 81~~~.
\end{equation}
According to the sum rules of Eq.  \eqref{branch1}, another way to get the same result is to consider the decomposition $SU(8) \supset SU(3) \times Sp(4)$,
and to either (i) sum the $SU(3)$ indices multiplied by the multiplicity factor counting the  number of times each occurs \big(given
by the dimension of the associated $Sp(4)$ representation\big), or to (ii) sum the $Sp(4)$ indices  multiplied by the multiplicity factor counting the number of times each occurs \big(given
by the dimension of the associated $SU(3)$ representation\big), both as read off from the $SU(8)$ branching expansions of Eqs. \eqref{decomp1}--\eqref{decompp2}, together with
\begin{equation}\label{decomp3}
63=(1,10)+(8,1)+(3,4)+(\overline{3},4)+(3,1)+(\overline{3},1)+(1,5)+(1,4)_1+(1,4)_2+(1,1)_1+(1,1)_2~~~.
\end{equation}
The needed index values for $SU(3)$ and
$Sp(4)$ are given in Tables II and III, and the needed multiplicities are given in Tables IV and V.

\begin{table} [h]
\caption{$SU(8)=A_7$ index values.}

\centering
\begin{tabular}{  c c c c c c}
\hline\hline
representation  & ~1~ & ~8~ & ~$\overline{28}$~ & ~56~ & ~63~ \\
index & ~0~ & ~1~ & ~6~ & ~15~ & ~16~  \\

\hline\hline

\end{tabular}
\label{one}
\end{table}

\begin{table} [h]
\caption{$SU(3)=A_2$ index values.}

\centering
\begin{tabular}{  c c c c }
\hline\hline
representation  & ~1~ & ~3,$\overline{3}$~ & ~8 ~\\
index & ~0~ & ~1~ & ~6~   \\

\hline\hline

\end{tabular}
\label{two}
\end{table}

\begin{table} [h]
\caption{$Sp(4)=C_2$ index values.}

\centering
\begin{tabular}{  c c c c c }
\hline\hline
representation  & ~1~ & ~4~ & ~5~ & ~10~  \\
index & ~0~ & ~1~ & ~2~ & ~6~  \\

\hline\hline

\end{tabular}
\label{three}
\end{table}

\vfill\eject

 \begin{table} []
\caption{$SU(3)$ representation multiplicities above the $SU(8)$ breaking scale.  The first column gives
the weight in $K$, the second the originating $SU(8)$ representation, and the next three columns
the multiplicities of the $SU(3)$ representations listed in the first row. The $K$ weights are the sums
of the weights in Eq. \eqref{beta} for representations of the same type, i.e., a vector 56 and a scalar 56,
and two $\overline{28}$. }

\centering
\begin{tabular}{  c c c c c c}
\hline\hline
$K$ weight &  $SU(8)$   & ~~1~~ &~~3,$\overline{3}$~~ & ~~8~~  \\
-26   & 8   &  5  &  1    & 0   \\
 -4  & $\overline{28}$   &  10  &   6   & 0  \\
 -3 & 56   &  11  &  15    &   0 \\
 11  & 63   & 25   &  10    &  1 \\

\hline\hline

\end{tabular}
\label{four}
\end{table}

\begin{table} [h]
\caption{$Sp(4)$ representation multiplicities above the $SU(8)$ breaking scale.  The first column gives
the weight in $K$, the second the originating $SU(8)$ representation, and the next four columns
the multiplicities of the $Sp(4)$ representations listed in the first row. The $K$ weights are the sums
of the weights in Eq. \eqref{beta} for representations of the same type, i.e., a vector 56 and a scalar 56,
and two $\overline{28}$.}

\centering
\begin{tabular}{  c c c c c c }
\hline\hline
$K$ weight  & $SU(8)$  & ~~1~~ & ~~4~~ & ~~5~~  &~~ 10~~\\
-26 & 8 & 4 & 1  & 0  & 0  \\
-4 & $\overline{28}$ & 7 & 4  & 1  & 0  \\
-3 & 56  & 8 &  7 & 4  & 0  \\
11 & 63 & 16 & 8  & 1  & 1  \\
\hline\hline

\end{tabular}
\label{five}
\end{table}

The calculation for (i) is (with the factor preceding each $\times$  the index and the factor following each $\times$  the multiplicity, and reading the rows in Table IV from bottom up),
\begin{equation}\label{icalc}
K=11(1 \times 10+6 \times 1)-3(1 \times 15) -4(1\times 6)-26(1\times 1)=81~~~,
\end{equation}
and the calculation for (ii) is (again with the factor preceding each $\times$  the index and the factor following each $\times$  the multiplicity, and reading the rows in Table V from bottom up),
\begin{equation}\label{iicalc}
K=11(1\times 8+2 \times 1 +6 \times 1)-3(1\times 7+2\times 4)-4(1\times 4+2 \times 1)- 26(1\times 1)=81~~~.
\end{equation}
The reason for repeating the unbroken $SU(8)$ calculation this way is that it anticipates the organizational method that we
shall use to get the beta functions after $SU(8)$ is broken.

Let us now do a similar analysis for the running couplings below the $SU(8)$ breaking scale.  Here only the $(8,1)$ and $(1,10)$ of the original $63$ vector
gauge bosons remain massless, and all of the $56$ scalars either get masses or become Goldstone modes that are absorbed into the massive vectors.  Turning to the
fermions, only the $56$ and one of the two $\overline{28}$ are present as complete multiplets; the other $\overline{28}$ is present only as the partial multiplet
given in Eq. \eqref{decomp2}.  As we have seen, the remaining parts of this $\overline{28}$ combine with the representation $8$ spin $\frac{3}{2}$ to give massive
spin $\frac{3}{2}$ states, which do not contribute to the running coupling below the $SU(8)$ breaking scale. (We will discuss shortly what happens when the spin $\frac{3}{2}$
mass scale is appreciably below the $SU(8)$ breaking scale.)  Thus, the analogs of Tables IV and V are as given in Tables VI
and VII,
with $\overline{28}_{\lambda_{\perp}}$ the complete multiplet and $\overline{28}_{\Delta\lambda}$ the partial multiplet,

 \begin{table} []
\caption{$SU(3)$ representation multiplicities below the $SU(8)$ breaking scale.  The first column gives
the weight in $K$, the second the originating $SU(8)$ representation, and the next three columns
the multiplicities of the $SU(3)$ representations listed in the first row.}

\centering
\begin{tabular}{  c c c c c c}
\hline\hline
$K$ weight & originating $SU(8)$   & ~~1~~ &~~3,$\overline{3}$~~ & ~~8~~  \\
 -2  & $\overline{28}_{\lambda_{\perp}}$   &  10  &   6   & 0  \\
 -2  & $\overline{28}_{\Delta \lambda}$   &  5  &   5   & 0  \\
 -2 & 56   &  11  &  15    &   0 \\
 11  & 63   & 0   &  0    &  1 \\

\hline\hline

\end{tabular}
\label{six}
\end{table}

\begin{table} []
\caption{$Sp(4)$ representation multiplicities below the $SU(8)$ breaking scale.  The first column gives
the weight in $K$, the second the originating $SU(8)$ representation, and the next four columns
the multiplicities of the $Sp(4)$ representations listed in the first row.}

\centering
\begin{tabular}{  c c c c c c }
\hline\hline
$K$ weight  & originating $SU(8)$  & ~~1~~ & ~~4~~ & ~~5~~  &~~ 10~~\\
-2 & $\overline{28}_{\lambda_{\perp}}$ & 7 & 4  & 1  & 0  \\
-2 & $\overline{28}_{\Delta\lambda}$ & 3 & 3  & 1  & 0  \\
-2 & 56  & 8 &  7 & 4  & 0  \\
11 & 63 & 0 & 0  & 0  & 1  \\
\hline\hline

\end{tabular}
\label{seven}
\end{table}

From Tables VI and VII, we get the analogs of the $K$ calculations of Eqs. \eqref{icalc} and \eqref{iicalc}.  For the $SU(3)$ running coupling,
we find from Table VI that the analog of Eq. \eqref{icalc} is
\begin{equation}\label{su3}
K=11(6 \times 1)-2(1 \times 15) -2(1\times 5)-2(1 \times 6)=14~~~,
\end{equation}
and for the $Sp(4)$ running coupling, we find from Table VII that the analog of Eq. \eqref{iicalc} is
\begin{equation}\label{sp4}
K=11(6 \times 1)-2(1 \times 7+2 \times 4)-2(1 \times 3+2\times 1) -2(1\times 4+2 \times 1)=14~~~;
\end{equation}
again, in these equations the factor preceding each $\times$ is the index and the factor following each $\times$ is  the multiplicity, and the rows in Tables VI and VII are read from
bottom up.  We see that despite the differing multiplicity and index values, the $SU(3)$ and $Sp(4)$ couplings run to low energies at
exactly the same rate.   For comparison, and as noted again later, for $SU(3)$ with six flavors of Dirac fermions (each equivalent to two Weyl fermions), the corresponding
calculation gives $K=11(6 \times 1) - 4( 1 \times 6)=42$, a faster rate of running than given by Eqs. \eqref{su3} and \eqref{sp4}.  The fact that $K$ remains positive after
$SU(8)$ breaking is a direct consequence of the BEH mechanism that gives a mass to the spin $\frac{3}{2}$ fermion and part of the $\overline{28}$ multiplets.  If these were not
removed from the calculation of Eqs. \eqref{su3} and \eqref{sp4}, the analogous calculation in both cases would give
\begin{equation}\label{negk}
K=11(6)-2(15)-4(6)-26(1) =-14~~~,
\end{equation}
and the running couplings would start to decrease, rather than increasing, below the $SU(8)$ breaking scale.  If the spin $\frac{3}{2}$ mass scale is significantly below the
$SU(8)$ breaking scale, the $K$ value of Eq. \eqref{negk} applies in the interval between them.

\section{Making contact with the Standard Model}

\subsection{Incorporating weak hypercharge}
Any unification of forces beyond  $SU(3)\times SU(2)\times U(1)$ of the standard model must explain the
pattern of $U(1)$ charges $Y$ shown in Table VIII, the so-called weak hypercharge values.  There are two
strategies in grand unification for explaining the pattern of $Y$ values.  The first, employed in the
$SU(5)$, $SO(10)$ and related models \cite{zee}, is to find a group for which the appropriately normalized
$U(1)$ generator, acting on the fermion representations, gives the standard model $Y$ values.   A variant of this strategy, termed ``flipped $SU(5)$'',
uses an extra $U(1)_X$, mixing with a $U(1)$ coming from $SU(5)$, to get standard model $Y$ values using
a modified weak isospin assignment of the fermions \cite{barr}.  The paper \cite{adler1} that introduced the $SU(8)$ model
studied here proposed to make contact
with flipped $SU(5)$, through a modulo 5 periodicity in the $U(1)$ charge structure, but this is not compatible
with the Coleman-Weinberg symmetry breaking pattern found in (I). Leaving aside the question of
getting the needed extra $U(1)_X$, the analysis of (I)  shows that $SU(8)$ breaks not to
$SU(3) \times SU(5)$, as needed for both standard and flipped $SU(5)$, but rather to the smaller
group $SU(3)\times Sp(4)$.

A second grand unification strategy for explaining the pattern of $Y$ values is employed in the $SU(3) \times SU(2)_L \times SU(2)_R \times
U(1)_{B-L}$ model \cite{mohapatra}  and related models such as the Pati-Salam model \cite{zee}.  It is based on the observation that the averages of $Y$ values for each
pair of consecutive rows in Table VIII, noted by $\overline{Y}$ in the final column of the table, is identical to
$\frac{1}{2}(B-L)$, with $B$ the baryon number and $L$ the lepton number.  The $Y$ values of the conjugated left-handed fermions are
all equal to $\pm \frac{1}{2}$ added to the average $\overline{Y}$, suggesting that there is an additional hidden $SU(2)$ group, with
third component generator $T_{3\,{\rm hidden}}= \pm \frac{1}{2}$ acting
on the conjugated left-handed fermions, that is  broken at an energy scale above the standard model regime.  The weak hypercharge is then
given by $Y=\overline{Y}+T_{3\,{\rm hidden}}$, and the electric charge is given as usual by $Q=T_3+Y$.

\begin{table} [h]
\caption{Fermions of a standard model family.  The first column gives the left-handed fermion name
(the conjugates of left-handed fermions are right-handed), the second column the color $SU(3)$ representation,
the third column the weak isospin $SU(2)$ representation, the fourth column the weak isospin value $T_3$,  the fifth column
the weak hypercharge $Y$, the sixth column the electric charge $Q$, and the seventh column the average $Y$, denoted by $\overline{Y}$,  for
each pair of two consecutive rows in the table, which is equal to half the difference between the baryon number $B$ and the
lepton number $L$.}
\centering
\begin{tabular}{  c c c c c c c}
&&&&&&\\
\hline\hline
~fermion~ & ~$SU(3)$~ & ~$SU(2)$~ & ~~$T_3$~~ & ~~$Y$~~ & ~~$Q=T_3+Y$~~ & ~~$\overline{Y}$=$\frac{1}{2}(B-L)$\\
\hline  \\
$u_L$ & 3 & 2 &$\frac{1}{2}$ &$\frac{1}{6}=\overline{Y}$ &$\frac{2}{3}$ &$\frac{1}{6}$\\
$d_L$ & 3 &2 &$-\frac{1}{2}$ & $\frac{1}{6}=\overline{Y}$&$-\frac{1}{3}$ &$\frac{1}{6}$\\
$d_L^*$ & $\overline{3}$ &1 &0 &$\frac{1}{3}=\overline{Y}+\frac{1}{2}$ &$\frac{1}{3}$ &$-\frac{1}{6}$ \\
$u_L^*$ & $\overline{3}$ &1 &0 &$-\frac{2}{3}=\overline{Y}-\frac{1}{2}$ &-$\frac{2}{3}$ &$-\frac{1}{6}$ \\
$\nu_L$ &1 &2 &$\frac{1}{2}$ &$-\frac{1}{2}=\overline{Y}$ &0 &$-\frac{1}{2}$\\
$\ell_L$ &1 &2 &-$\frac{1}{2}$ &$-\frac{1}{2}=\overline{Y}$ &$-1$ &$-\frac{1}{2}$\\
$\ell_L^*$&1 &1 &0 &$1=\overline{Y}+\frac{1}{2}$ &1 &$\frac{1}{2}$\\
$\nu_L^*$&1 &1 & 0&$0 =\overline{Y}-\frac{1}{2}$&0 &$\frac{1}{2}$\\
\hline\hline

\end{tabular}
\label{eight}
\end{table}

There are two hints
that this strategy is the correct one for the $SU(8)$ theory broken initially by the Coleman-Weinberg mechanism.  The first hint is that
the group $Sp(4)$ admits the symmetry breaking pattern $Sp(4)\supset SU(2) \times SU(2)$ as needed to get the non-Abelian part of the
second strategy models.
Although there is no additional massless $U(1)$ after $SU(8)$ breaking, the second hint comes from examining the structure of the
residual vector meson mass spectrum after $SU(8)$ breaking, as given in Table IX.  We see that the vector meson with the lowest nonzero
mass matrix eigenvalue, $0.0874$, and that with the highest mass matrix eigenvalue, $1.384$, are both $U(1)$ generators that commute
with the $SU(3)$ and $Sp(4)$ generators.  The corresponding generators acting on the fundamental $8$ representation, with
components numbered sequentially 1 through 8 and normalized so that
the 2,3,7, and 8 components are $-1$, are
\begin{align}\label{gencomp}
{\rm eigenvalue}=0.0874  \leftrightarrow& {\rm generator} =(2.97,-1.00,-1.00,0.343,0.343,0.343,-1.00,-1.00)~~~,\cr
{\rm eigenvalue} =1.384  \leftrightarrow& {\rm generator} =(-2.04,-1.00,-1.00,2.01,2.01,2.01,-1.00,-1.00)~~~.\cr
\end{align}
We see that these are within a few percent of the trace orthogonal generators
\begin{align}\label{gencomp1}
{\rm eigenvalue}=0.0874  \leftrightarrow& {\rm generator} \simeq(3,-1,-1,\frac{1}{3},\frac{1}{3},\frac{1}{3},-1,-1)~~~,\cr
{\rm eigenvalue} =1.384  \leftrightarrow& {\rm generator} \simeq(-2,-1,-1,2,2,2,-1,-1)~~~.\cr
\end{align}
Interestingly, the  first line in Eq. \eqref{gencomp1}, corresponding to the smallest vector meson mass eigenvalue,  has a ratio of color $SU(3)$ charges to $Sp(4)$ charges equal to $-\frac{1}{3}$, which is the same as
the ratio of the $\overline{Y}=\frac{1}{2}(B-L)$ values of quarks to leptons in the final column of Table VIII.  This is again suggestive that
the $SU(8)$ model may make contact with the observed quarks and leptons through an analog of the second strategy models based on $SU(3) \times SU(2) \times SU(2)
\times U(1)_{B-L}$, and this is the strategy that we will pursue in what follows.

\begin{table} [h]
\caption{Algebraic formulas expressed in terms of  $a, \, b$, and numerical values at the minimum $a=0.59762...$, $b=0.67199...$, for the vector gauge field eigenvalues $m_i^2$, \,i =1,...,63,  as well as the number of degenerate eigenvalues of each type, and the $SU(3) \times Sp(4)$ representation content.  In the final line, $E=-\frac{1}{8}(14a^2+15b^2)$ and $F=\frac{3}{4}a^2b^2$,
as given in Eq. (13) of (I).}

\centering
\begin{tabular}{ c c c c}
\hline\hline
eigenvalue degeneracy  & algebraic formula&  numerical value &~~~ $SU(3) \times Sp(4)$  content\\
\hline
  18  &  0 & 0.000&$(1,10)+(8,1)$\\
   8  &  $\frac{1}{2}a^2$ & 0.179 &$(1,4)_1+(1,4)_2$\\\
   24 & $\frac{1}{2} (a^2+b^2) $ & 0.404 &$(3,4)+(\overline{3},4)$\\
   6  &  $\frac{1}{2} (2a^2+b^2)$ & 0.583 &$(3,1)+(\overline{3},1)$\\
   5  & $2a^2$  &0.714  &$(1,5)$\\
   1,\,1  & {2~\rm roots $x_{1,2}$ of} ~~$x^2+Ex+F=0$  &0.0874,\,1.384 &$(1,1)_1+(1,1)_2$ \\
   \hline\hline

\end{tabular}
\label{abalgebraic}
\end{table}
\subsection{Three conjectures}

Motivated by the observations in the preceding section, we now make three conjectures about the behavior of the
$SU(8)$ theory when full radiative corrections are included.
\begin{itemize}

\item  First Conjecture.  As the energy decreases, the Coleman-Weinberg potential evolves in such a way
that at a scale $M_{Sp(4)}$ the symmetry $c=a$ (with $a=\overline{\phi}^{[123]}$ and $c=\overline{\phi}^{[178]}$)  is broken,\footnote{See Sec. III
of (I) for a detailed discussion of the $c=a$ symmetry of the Coleman-Weinberg potential minimum.} corresponding to the breaking of $Sp(4)$ symmetry according to
$Sp(4) \supset SU(2) \times SU(2)$. This can happen if the coefficient $A$ of the $\epsilon^2=\big((c-a)/2\big)^2$ term
in the expansion of the potential $V$ in Eq. (12) of (I)  evolves from positive to negative, changing the
potential from one with a stable minimum at $c=a$ to a ``Mexican hat'' potential with an unstable extremum at $c=a$,
and two stable minima at $|c-a|>0$.
Since $a$ and $c$ are
associated with the two distinct $SU(2)$ subgroups, the theory will break asymmetrically, rather than
enforcing symmetry between the two $SU(2)$ subgroups as in left-right symmetric models.  To see that group theory permits
this breaking, we note from Table XII that $a-c=\overline{\phi}^{[123]}-\overline{\phi}^{[178]}$ lies in the $SU(3)\times Sp(4)$
representation $(1,5)$, and that under $Sp(4) \supset SU(2) \times SU(2)$ the 5 of $Sp(4)$ decomposes into representations of
$SU(2) \times SU(2)$ as $5=(1,1)+(2,2)$.  With the nonzero expectation $a-c$ lying in the $(1,1)$, the $(2,2)$ will be massless
Goldstone bosons, with the correct quantum numbers to be absorbed into the 4 broken generators of $Sp(4)$, giving them
masses by the BEH mechanism.

\item   Second Conjecture.  As the energy decreases further from $M_{Sp(4)}$, the theory runs to
an infrared fixed point starting at an energy scale $M_U$ at which the mass matrix eigenvalue $0.0874$ becomes exactly zero, and the
corresponding generator becomes that of the first line of Eq. \eqref{gencomp1}.\footnote{A priori, $M_U$ could be above or below
$M_{Sp(4)}$, but the coupling constant analysis given below suggests  that $M_U$  lies below $M_{Sp(4)}$.}  Rescaling by a factor
of 2 so that the 2,3,7 and 8 components are $-\frac{1}{2}$, this zero eigenvalue of the gauge field mass
matrix is then associated with
$U(1)$ generators $G$ acting on the fermions in the 8 and 56 representations, and $-G$ acting on the
fermions in the $\overline{28}$ representation, with $G$ given by
\begin{equation}\label{gencomp3}
G=(\frac{3}{2}, -\frac{1}{2},-\frac{1}{2},\frac{1}{6},\frac{1}{6},\frac{1}{6},-\frac{1}{2},-\frac{1}{2})~~~.
\end{equation}
Evidently, $G$ will play the role of a proto-$U(1)_{\frac{1}{2}(B-L)}$ generator.  An interesting aspect of the $U(1)$ generator
$G$ of Eq. \eqref{gencomp3} is that the components $\overline{\phi}^{[123]}$, $\overline{\phi}^{[456]}$, and $\overline{\phi}^{[178]}$ of $\phi$ at the symmetry-breaking minimum of the Coleman-Weinberg potential all have the same $G$ value  of $\frac{1}{2}$.\footnote{This is also the same as the $G$ value of the Higgs boson, suggesting that a vacuum tadpole can be viewed as a ``spurion'' that absorbs or emits $G=\frac{1}{2}$.}
An equivalent observation is that the most general $U(1)$ generator that commutes with the $SU(3)$ and $Sp(4)$ generators has the form
$(c,a,a,b,b,b,a,a)$ with $c+4a+3b=0$ to give trace zero. Requiring that $\overline{\phi}^{[123]}$, $\overline{\phi}^{[456]}$, and $\overline{\phi}^{[178]}$ all have the same value of this generator gives the additional condition $c+2a=3b$, requiring $a=-3b,\, c=9b$, which up to normalization is Eq. \eqref{gencomp3}.\footnote{This mechanism for emergence of an unbroken $U(1)$ suggests that cosmological monopoles \cite{preskill} will not form.  In the initial stage of symmetry breaking $SU(8)\supset SU(3)\times Sp(4)$, arising from the 56 scalar field, the
groups $SU(8)$, $SU(3)$ and $Sp(4)$ that are involved are all simply connected, and thus \cite{weinberghomo}  $\pi _2\big(SU(8)/(SU(3)\times Sp(4))\big)=\pi_1\big(SU(3)\times Sp(4)\big)=0$, and no monopoles are formed.  Since no scalar field components survive the initial stages of symmetry breaking, when an unbroken $U(1)$ emerges at lower energies,  there is no long range scalar field available for monopole formation.}

\item Third Conjecture.   We have seen that before
$Sp(4)$ breaking the running couplings of the $SU(3)$ and $Sp(4)$ subgroups evolve at the same rate, with $K=14$.  The natural co-running of these couplings is a prerequisite for  implementing dynamical symmetry breaking, as  emphasized early on by
Weinberg \cite{weinberg1}.  After the asymmetric breaking of $Sp(4)$ the $K$ value of $SU(3)$ will remain the same, but the $K$
values of the two new $SU(2)$ subgroups will no longer be 14.
Since the index of the adjoint 3 representation of $SU(2)$ is 4, the gluon contribution to $K$
for the $SU(2)$ subgroups will change from the $11(6 \times 1)$ of Eqs. \eqref{su3} and \eqref{sp4} to $11(4 \times 1)$, with the
fermion contribution unchanged by virtue of the index sum rule of Eq. \eqref{branch1}.  Thus before taking effects of the asymmetric
breaking into account, the nominal $K$ value for each of the two $SU(2)$ subgroups will be $K=-8$.
Taking into account the asymmetric breaking of $Sp(4)$, we postulate that one $SU(2)$ subgroup will have a $K$ value
higher than $-8$ given by $K=-8+\Delta_+$,  and one will have a $K$ value lower than $-8$ given by $K=-8-\Delta_-$, with
 $\Delta_{\pm} \geq 0$.  We call the $SU(2)$ that evolves with
the higher value of $K$ the ``technicolor group'' $SU(2)_{TC}$, and assume that it leads to binding of the original
fermions at a scale $M_{TC}$ above the standard model electroweak scale.  We call the $SU(2)$ that evolves with
the lower value of $K$ the ``electroweak group'' $SU(2)_{EW}$, and associate it with the electroweak component of the standard
model.  Since it is evolving at a slower rate than the $SU(3)$ group, it can remain weak when the ``color''
interactions associated with the $SU(3)$ group become strong.\footnote{Our conjectures are related to ideas long in the
literature under the names of ``technicolor, ``extended technicolor'', ``walking technicolor'' and ``little higgs''.  For seminal papers,
see \cite{susskind} and \cite{weinberg1}, \cite{weinberg2}.  For good reviews see  \cite{lane}, \cite{piai},  \cite{schmaltz} and \cite{perelstein}.  See also \cite{raby} on ``tumbling''.}

\end{itemize}

\subsection{Constraints coming from extrapolation of the standard model couplings}

We turn now to an analysis of the constraints on the postulated energy scales $M_{Sp(4)}$, $M_{TC}$, and $M_U$, and on the postulated
running coupling evolution asymmetries $\Delta_{\pm}$, implied by extrapolation from the standard model couplings at the $Z$ meson mass
$M_Z=91.2\,{\rm GeV}=0.0912\,{\rm TeV}$.  Letting $g_s$ denote the strong coupling, $g$ the $SU(2)_{EW}$ coupling, and $g^{\prime}$
the $U(1)$ coupling of the standard model \cite{weinberghomo}, we have
\begin{align}\label{smcouplings}
g_s^2(M_Z)=&1.485~~~,\cr
g^2(M_Z)=&0.4246~~~,\cr
g^{\prime\,2}(M_Z)=&0.1278~~~.\cr
\end{align}
These couplings can be evolved up in energy to the postulated scale $M_{TC}$ at which we conjecture that preons bind to form
the standard model fermions, using the customary evolution equations \cite{weinberghomo}
\begin{align}\label{evolution}
\frac{1}{g_s^2(M_{TC})}=&\frac{1}{g_s^2(M_Z)}-\frac{7}{8\pi^2} \log\left(\frac{M_Z}{M_{TC}}\right)~~~,\cr
\frac{1}{g^2(M_{TC})}=&\frac{1}{g^2(M_Z)}-\frac{10}{24\pi^2} \log\left(\frac{M_Z}{M_{TC}}\right)~~~,\cr
\frac{1}{g^{\prime\,2}(M_{TC})}=&\frac{1}{g^{\prime\,2}(M_Z)}+\frac{20}{24\pi^2} \log\left(\frac{M_Z}{M_{TC}}\right)~~~.\cr
\end{align}

We focus first on the non-Abelian couplings $g_s^2$, $g^2_{EW}\equiv g^2$, and $g^2_{TC}$  for the
 $SU(3)$,  $SU(2)_{EW}$, and $SU(2)_{TC}$  groups that act between the energy scales $M_{TC}$ and $M_{Sp(4)}$.  Their evolution is given by
the formulas
\begin{align}\label{evolution1}
\frac{1}{g_s^2(M_{Sp(4)})}=&\frac{1}{g_s^2(M_{TC})}-\frac{7}{24\pi^2} \log\left(\frac{M_{TC}}{M_{Sp(4)}}\right)~~~,\cr
\frac{1}{g^2_{EW}(M_{Sp(4)})}=&\frac{1}{g^2_{EW}(M_{TC})}+\frac{8+\Delta_-}{48\pi^2} \log\left(\frac{M_{TC}}{M_{Sp(4)}}\right)~~~,\cr
\frac{1}{g^2_{TC}(M_{Sp(4)})}=&\frac{1}{g^2_{TC}(M_{TC})}+\frac{8-\Delta_+}{48\pi^2} \log\left(\frac{M_{TC}}{M_{Sp(4)}}\right)~~~.\cr
\end{align}
Combining the first two lines of Eq. \eqref{evolution} with the corresponding lines of Eq. \eqref{evolution1} we get the equations for
evolution of $g_s^2$ and $g^2_{EW}$ between $M_Z$ and $M_{Sp(4)}$,
\begin{align}\label{evoution3}
\frac{1}{g_s^2(M_{Sp(4)})}=&\frac{1}{g_s^2(M_Z)}-\frac{7}{24\pi^2} \log\left(\frac{M_{TC}}{M_{Sp(4)}}\right)
-\frac{7}{8\pi^2} \log\left(\frac{M_Z}{M_{TC}}\right)~~~,\cr
\frac{1}{g^2_{EW}(M_{Sp(4)})}=&\frac{1}{g^2(M_Z)}+\frac{8+\Delta_-}{48\pi^2} \log\left(\frac{M_{TC}}{M_{Sp(4)}}\right)
-\frac{10}{24\pi^2} \log\left(\frac{M_Z}{M_{TC}}\right)~~~.\cr
\end{align}

Since the $SU(3)$ and $Sp(4)$ couplings co-evolve below the $SU(8)$ breaking scale, and assuming that there are no finite coupling renormalizations at the $Sp(4)$ breaking scale, the couplings $g_s^2(M_{Sp(4)})$, $g^2_{EW}(M_{Sp(4)})$, and $g^2_{TC}(M_{Sp(4)})$ will be equal to a common
value.  Also, since we are defining $M_{TC}$ as the scale at which the technicolor group $SU(2)_{TC}$ becomes strongly coupled, we
can effectively assume $\frac{1}{g^2_{TC}(M_{TC})}=0$.  This give two conditions on the four parameters $M_{TC}$, $M_{Sp(4)}$, $\Delta_+$,
and $\Delta_-$, which can be used to generate the values for these parameters given in Table X.  We see that a large value of $\Delta_+$
is needed in order for the technicolor interaction to become strong before the strong interaction, and that the value of $\Delta_-$ required
to match the $SU(2)_{EW}$ coupling at $M_Z$ is in general considerably smaller than $\Delta_+$.

We consider next the emergent $U(1)$ coupled to $G$, that is postulated in the second conjecture to emerge at a scale $M_U$.  Its running
coupling $g_U$ will obey
\begin{equation}\label{urunning1}
\mu \frac{dg_U}{d\mu}=\frac{g_U^3}{24\pi^2} \sum_i G_i^2~~~,
\end{equation}
with $\sum_i G_i^2=84.667$ a sum of the $G^2$ values for all of the individual tensor components listed in Tables XI and XII.
Integrating this, we get
\begin{equation}\label{urunning2}
\frac{1}{g_U^2(M_U)}=\frac{1}{g_U^2(M_{TC})}+\frac{84.667}{12\pi^2} \log\left(\frac{M_{TC}}{M_U}\right)~~~.
\end{equation}
Between $M_{TC}$ and $M_Z$, the $U(1)$ running coupling for the weak hypercharge $Y$  evolves according to
the third line of Eq. \eqref{evolution}.  At $M_{TC}$ we must match the coupling $g^{\prime}$ to $g_{TC}$ and $g_U$, to reflect
the fact that $Y=\overline{Y}+T_{3\,{\rm hidden}}$, with $\overline{Y}=G$ and $T_{3\,{\rm hidden}}=T_{3TC}$.  The matching relation,
derived in the Appendix A, is
\begin{equation}\label{matching}
\frac{1}{g^{\prime\,2}}=\frac{1}{g_{TC}^2}+\frac{1}{g_U^2},
\end{equation}
which when $1/g_{TC}^2=0$ reduces to  $g^{\prime}=g_U$.
Combining this with Eq. \eqref{urunning2} and with the third line of  Eq. \eqref{evolution}, and noting that
$0\leq \frac{1}{g_U^2(M_U)}$, we get the inequality
\begin{equation}\label{inequality}
0\leq\frac{1}{g^{\prime\,2}(M_Z)}+ \frac{84.667}{12\pi^2} \log\left(\frac{M_{TC}}{M_U}\right)+  \frac{20}{24\pi^2} \log\left(\frac{M_Z}{M_{TC}}\right)~~~.
\end{equation}
This can be rearranged into the form
\begin{equation}\label{inequality1}
\log\left(\frac{M_U}{M_Z}\right)\leq \frac{12\pi^2}{84.667 g^{\prime\,2}(M_Z)}+ 0.882 \log\left(\frac{M_{TC}}{M_Z}\right)~~~,
\end{equation}
which was used to compute the upper bounds on $M_U$ given in the final column of Table X.  We see that for large values of $M_{Sp(4)}$,
$M_U$ is considerably below $M_{Sp(4)}$.

\begin{table} [ht]
\caption{Values of $\Delta_+$, $\Delta_-$, and the upper bound $M_U$, versus $M_{TC}$ and $M_{Sp(4)}
$.}

\centering
\begin{tabular}{ c c c c c}
~~$M_{Sp(4)}(TeV)$~~ &~~$M_{TC}(TeV)$~~&~~$\Delta_+$~~&~~$\Delta_-$~~&~~$M_U(TeV)$~~\\
\hline\hline
$10^5$ & 10 & 78 & 53 & $3.3\times 10^5$ \\
 $10^5$ & 100&111&71& $2.5\times 10^6$ \\
 $10^5$ & 1000&176&107&$1.9\times 10^7$\\
\hline
$10^9$ & 10 & 50 & 16 & $3.3\times 10^5$ \\
 $10^9$ & 100&60&18& $2.5\times 10^6$ \\
 $10^9$ & 1000&73&21&$1.9\times 10^7$\\
\hline
 $10^{13}$ & 10 & 41 & 3.1& $3.3\times 10^5$ \\
 $10^{13}$ & 100&46&3.4& $2.5\times 10^6$ \\
 $10^{13}$ & 1000&53&3.7&$1.9\times 10^7$\\
\hline

\hline\hline

\end{tabular}
\label{lparam}
\end{table}

\subsection{$U(1)$ generator values for the $56$ and $\overline{28}$  fermions below the $SU(8)$ breaking scale}

According to our second conjecture, the theory runs towards an infrared fixed point at and below which there is a massless $U(1)$ generator $G$
given by Eq. \eqref{gencomp3}.  Applying $G$ to the states in the $56$ representation given in Table VII of (I), and $-G$ to the
states in the $\overline{28}$ representation given in Table VI of (I), we get the list of $U(1)$ quantum number assignments
given in Tables XI and XII.

\begin{table} [ht]
\caption{Representation content and postulated $U(1)$ generator $-G$ values of $SU(8)$ $\overline{28}$.
All of these states appear in $\lambda_{\perp}$, while the states marked with a superscript D (signifying doublet) also appear
in $\Delta \lambda$.}

\centering
\begin{tabular}{ c c c c}
\hline\hline
~~~$SU(2) \times SU(3) \times SU(2)$ ~~~  &~~~$SU(3) \times Sp(4)$~~~ &~~~ tensor components &  ~~~$-G$ value\\
\hline
 (1,1,1)   & (1,1)  & [23]+[78] & 1 \\
 (1,3,1)   &(3,1)  & [45],\,[46],\,[56] & $-\frac{1}{3}$ \\
(1,$\overline{3}$,1)$^D$    & ($\overline{3}$,1)  &[14],\,[15],\,[16] &$ -\frac{5}{3}$ \\
(2,1,1)    &(1,4)   & [12],\,[13] &  -1\\
 (1,1,2)   &(1,4)   & [17],\,[18] & -1 \\
 (1,1,1)$^D$   &(1,5)   &[23]$-$[78] & 1 \\
 (2,1,2)$^D$   &(1,5)   & [27],\,[28],\,[37],\,[38] & 1 \\
(2,$\overline{3}$,1)$^D$    & ($\overline{3}$,4)  & [24],\,[25],[26],\,[34],\,[35],\,[36] & $\frac{1}{3}$ \\
(1,$\overline{3}$,2)$^D$    & ($\overline{3}$,4)  & [47],\,[48],\,[57],\,[58],\,[67],\,[68] & $\frac{1}{3} $\\
\hline\hline

\end{tabular}
\label{28cont}
\end{table}

\begin{table} [ht]
\caption{Representation content and postulated $U(1)$ generator $G$ values of the $SU(8)$ 56 states $\chi$.}

\centering
\begin{tabular}{ c c c c}
\hline\hline
~~~$SU(2) \times SU(3) \times SU(2)$ ~~~  & ~~~$SU(3) \times Sp(4)$~~~&~~~ tensor components& ~~~$G$ value  \\
\hline
(1,1,1)    & (1,1)  & [123]+[178],\,[456] &$ \frac{1}{2},\frac{1}{2}$\\
 (1,1,2)   & (1,4)  &[237],\,[238] & $-\frac{3}{2}$\\
 (2,1,1)   & (1,4)  & [278],\,[378] & $-\frac{3}{2}$\\
 (1,1,1)   & (1,5)  & [123]$-$[178] &$ \frac{1}{2}$\\
 (2,1,2)   & (1,5)  &[127],\,[128],\,[137],\,[138] & $\frac{1}{2}$\\
 (1,3,1)   & (3,1)  & [234]+[478],\,[235]+[578],\,[236]+[678] & $ -\frac{5}{6}$\\
(1,$\overline{3}$,1)    & ($\overline{3}$,1)  &[145],\,[146],\,[156] & $\frac{11}{6}$ \\
(2,3,1)    &(3,4)   & [124],\,[125],\,[126],\,[134],\,[135],\,[136] &$ \frac{7}{6}$\\
(1,3,2)    &(3,4)   & [147],\,[148],\,[157],\,[158],\,[167],\,[168] & $\frac{7}{6}$\\
(2,$\overline{3}$,1)    & ($\overline{3}$,4)  &[245],\,[246],\,[256],\,[345],\,[346],\,[356] & $-\frac{1}{6}$\\
 (1,$\overline{3}$,2)   &($\overline{3}$,4)   & [457],\,[458],\,[467],\,[468],\,[567],\,[568] &$-\frac{1}{6}$\\
(1,3,1)    & (3,5)    &  [234]$-$[478],\,[235]$-$[578],\,[236]$-$[678]   & $-\frac{5}{6}$\\
  (2,3,2)  &  (3,5) &   [247],\,[248],\,[257],\,[258],\,[267],\,[268] & $-\frac{5}{6}$\\
  ~~&~~& [347],\,[348],\,[357],\,[358],\,[367],\,[368] &~~\\
\hline\hline

\end{tabular}
\label{56cont}
\end{table}

We see immediately from these tables that the fermions of the 56 and $\overline{28}$ representations cannot
give all of the standard model fermions of Table VIII.  For example, $U(1)$ generator values $\frac{1}{2}$ and $-\frac{1}{6}$ appear
in Tables XI and XII, but $U(1)$ generator values $-\frac{1}{2}$ and $\frac{1}{6}$ are missing (or vice-versa if we reverse the overall sign
of the $U(1)$ generator $G$.)  Thus, in order to make contact with the fermions of the standard model, we must examine the possibility that the 56
and $\overline{28}$ fermions are preons, from which the standard model fermions are formed as composites.\footnote{For a good
introduction to composite quarks and leptons, see Peskin \cite{peskin}.  For more recent reviews of preon models, see Fritzsch \cite{fritzsch}
and Wesenberg \cite{wesen}. }

\subsection{Synopsis of three fermion and five fermion candidates for composite leptons and quarks}
To study whether the $SU(8)$ model fermions can be preons, we did a computer search of the quantum numbers of
all three fermion and five fermion combinations, using the enumeration of $\overline{28}$ and 56 representation states
in Tables XI and XII. Sample results for the numbers of quark and lepton candidates are given in Tables XIII and XIV, in which four possible cases were computed:
$\overline{Y}= \xi \sum_{{\rm preons}~i} G_i$ with $\xi=1$ or $\xi=-1$, and interchange (or non-interchange) of $SU(3)$  3 and
$\overline{3}$.  These cases arise because the sign chosen for the $U(1)$ generator $G$, and the labeling of quarks as $3$ rather than $\overline{3}$,
are both arbitrary conventions.  In computing
Tables XIII and XIV, we required that  at most one preon be an $SU(2)_{TC}$ singlet, and that the $G$ value of this preon times the
sum of the $G$ values of the other preons be negative, as suggested by the demand of a net attractive force.  When the latter requirement is omitted, the
nonzero entries in Tables XIII and XIV increase as expected, but the 0 entries all remain unchanged.  When this requirement is
strengthened to require that no preons be  $SU(2)_{TC}$ singlets, additional 0 entries appear in the $SU(2)_{TC}$ singlet rows, since an
odd number of doublets cannot combine to give a singlet, and the entries in the $SU(2)_{TC}$ doublet rows are unchanged.

From Tables XIII and XIV, we see that the $SU(8)$ model can potentially account for all leptons, but not quarks, as 3 preon composites, and can potentially account for all quarks, but not all leptons, as 5 preon composites, using either the cases $\xi=1$ with no
$3\leftrightarrow\overline{3}$ (fifth column), or $\xi=-1$ with $3\leftrightarrow\overline{3}$ (eighth column).

\subsection{Synopsis of two fermion candidates for a composite Higgs boson}

Scalar composites formed from the left chiral 56 and $\overline{28}$ fields are studied  in Appendix B of \cite{adler1} using real Majorana
representation $\gamma$ matrices.  For any two
left chiral fields $\Psi_{L1}$ and $\Psi_{L2}$, the scalar $\overline{\Psi_{L1}} \Psi_{L2}=0$, but the scalar $\overline{\Psi^c_{L1}} \Psi_{L2}\neq 0$,  with
$c$ denoting charge conjugation. Since $\overline{\Psi^c_{L}} =\Psi_L^T i\gamma^0$ involves the Dirac transpose $T$ but no complex conjugation, the group representation
content of $\overline{\Psi^c_{L1}} \Psi_{L2}$ is simply the direct product of the representation content of $\Psi_1$ and $\Psi_2$.  So we can
get the quantum numbers of two preon scalar composites\footnote{To check that the composite $S\equiv\Psi_{L1}^Ti\gamma^0\Psi_{L2}$ is a two preon state, and not a preon-antipreon
state, form the commutator with the number operator $N=\int d^3x (\Psi_{L1}^{\dagger}\Psi_{L1}+\Psi_{L2}^{\dagger}\Psi_{L2})$.  Using the
symmetry $\Psi_{L1}^Ti\gamma^0\Psi_{L2}=\Psi_{L2}^Ti\gamma^0\Psi_{L1}$, it is easy to verify that the two commutator terms add
rather than subtract, giving $[S,N]= 2S$, corresponding to a two preon state.}
by taking direct products of the  representations, and  sums of the postulated $U(1)$ generator values, given in Tables XI and XII.

The standard model Higgs field is an $SU(2)_{EW}$ doublet.  If it is an $SU(2)_{TC}$ singlet it must have weak hypercharge $Y=\overline{Y}=\frac{1}{2}$ (so its complex conjugate has $Y=\overline{Y}=-\frac{1}{2}$), while if it is an $SU(2)_{TC}$ doublet, making
it a bi-doublet Higgs as assumed in left-right symmetric models, it must have $\overline{Y}=0$.  In Table XV we enumerate Higgs candidates
corresponding to these three choices of quantum numbers.  The computation was done with the restriction that both preons in the composite should be
$SU(2)_{TC}$ doublets; when this restriction is dropped, the nonzero entries 5 increase to 18, but the 0 entries all remain unchanged.
For all 5 candidates in Table XV $G_1G_2<0$, and so the $U(1)$ force, as well as the $SU(2)_{TC}$ force,  is attractive.
We see that the $SU(8)$ model can potentially account for an $SU(2)_{TC}$ singlet Higgs, but not for a bi-doublet Higgs, using either the cases
$\xi=1$ or $\xi=-1$.

\begin{table} [h]
\caption{Columns 5 through 8 give the number of candidates for three fermion composites in the $SU(3)$, $SU(2)_{EW}$ and
$SU(2)_{TC}$ representations listed in columns 1 through 3,  with the $\overline{Y}=\xi (G_1+G_2+G_3)$ value listed in column 4.
$G_i$ is the $-G$ value for a constituent fermion in the $\overline{28}$ as listed in Table XI, or the $G$ value for a constituent fermion in the 56 as listed in Table XII.}

\centering
\begin{tabular}{  c c c c c c c c}
\hline\hline

~~~$SU(3)$~~~ & ~~~$SU(2)_{EW}$~~~ & ~~~$SU(2)_{TC}$~~~& ~~~$\overline{Y}$~~~& ~~~$\xi=1$~~~&  ~~~ $\xi=1$ ~~~&  ~~~$\xi=-1$~~~& ~~~ $\xi=-1$~~~\\
        &              &             &           &        & ~$3\leftrightarrow\overline{3}$ &    &~$3\leftrightarrow\overline{3}$\\
\hline
 3 & 2& 1& $\frac{1}{6}$& 0&0 &0 & 0\\
$\overline{3}$  &1 &2 &$-\frac{1}{6}$ &0 &0 &0 &0 \\
 1 & 2& 1&$-\frac{1}{2}$ &15 &15 &43 & 43\\
 1 &1 & 2&$\frac{1}{2}$ &14 &14 &5 &5 \\

\hline\hline

\end{tabular}
\label{nine}
\end{table}

\begin{table} [h]
\caption{Columns 5 through 8 give the number of candidates for five fermion composites in the $SU(3)$, $SU(2)_{EW}$ and
$SU(2)_{TC}$ representations listed in columns 1 through 3,  with the $\overline{Y}=\xi (G_1+G_2+G_3+G_4+G_5)$ value listed in column 4.  $G_i$ is the $-G$ value for a constituent fermion in the $\overline{28}$ as listed in Table XI, or the $G$ value for a constituent fermion in the 56 as listed in Table XII.}
\centering
\begin{tabular}{  c c c c c c c c}
\hline\hline

~~~$SU(3)$~~~ & ~~~$SU(2)_{EW}$~~~ & ~~~$SU(2)_{TC}$~~~& ~~~$\overline{Y}$~~~& ~~~$\xi=1$~~~&  ~~~ $\xi=1$ ~~~&  ~~~$\xi=-1$~~~& ~~~ $\xi=-1$~~~\\
        &              &             &           &        & ~$3\leftrightarrow\overline{3}$ &    &~$3\leftrightarrow\overline{3}$\\
\hline
 3 & 2& 1& $\frac{1}{6}$&418 &0 &0 &211 \\
$\overline{3}$  &1 &2 &$-\frac{1}{6}$ &31 &0 &0 &61 \\
 1 & 2& 1&$-\frac{1}{2}$ &0 &0 & 37& 37\\
 1 &1 & 2&$\frac{1}{2}$ & 11& 11&0 &0 \\

\hline\hline

\end{tabular}
\label{ten}
\end{table}

\begin{table} [h]
\caption{Columns 5 and 6 give the number of candidates for two fermion scalar  composites in the $SU(3)$, $SU(2)_{EW}$ and
$SU(2)_{TC}$ representations listed in columns 1 through 3,  with the $\overline{Y}=\xi (G_1+G_2)$ value listed in column 4.
$G_i$ is the $-G$ value for a constituent fermion in the $\overline{28}$ as listed in Table XI, or the $G$ value for a constituent fermion in the 56 as listed in Table XII.  }

\centering
\begin{tabular}{  c c c c c c }
\hline\hline

~~~$SU(3)$~~~ & ~~~$SU(2)_{EW}$~~~ & ~~~$SU(2)_{TC}$~~~& ~~~$\overline{Y}$~~~& ~~~$\xi=1$~~~&  ~~~ $\xi=-1$ ~~~~~~\\
\hline
 1 & 2& 1& $\frac{1}{2}$& 0&5\\
1  &2 &1 &$-\frac{1}{2}$ &5 & 0 \\
 1 & 2& 2&$0$ &0 &0\\

\hline\hline

\end{tabular}
\label{eleven}
\end{table}

\subsection{A mechanism for a doublet-singlet three family structure}

There are enough candidates for composite leptons and quarks in Tables XIII and XIV to give three lepton plus quark families with different
internal structures, but without family triplet symmetry relations among the families. If one interprets the observed three families as indicating not a triplet structure, but instead a family doublet (perhaps the lightest two families) plus an extra singlet (the heaviest family), there is a preonic construction within our model that gives this, as follows.  We note from Table XI that representations marked with a superscript D
appear in both $\Delta \lambda$ and $\lambda_{\perp}$ and so are preonic family doublets, giving four potential contributors to composite doublets if the preonic doublet  $(1,1,1)$ is excluded.
Also, from Table XII we see that the representation $(1,3,1)(G=-\frac{5}{6})$ appears twice, giving a fifth potential contributor to a composite doublet.
We reduce this list of five to four by assuming that a preonic doublet which can bind to form a Higgs candidate in Table XV is split, with one
member binding into the Higgs and one member contributing to quarks and leptons as a preonic singlet, and we exclude from the preonic singlets list
the other preon that binds with this split doublet to form the Higgs.   Specifically, corresponding to two  $\xi=1$ cases in Table XV
with a family doublet preon binding to a singlet preon,
either we treat
$(2,1,2)(G=1)$ from Table XI as a split doublet, and $(2,1,1)(G=-\frac{3}{2})$ from Table XII as an excluded singlet, designating this Case (1),
or we treat $(2,\overline{3},1)(G=\frac{1}{3})$ from Table XI as a split doublet, and $(2,3,2)(G=-\frac{5}{6})$ from Table XII
as an excluded singlet, designating this Case (2). Then requiring quark doublets to have one family doublet preon plus four singlet preons, lepton
doublets to have one family doublet preon plus two singlet preons, and quark and lepton singlets to respectively be constructed entirely from singlet
preons, we get the enumeration of candidates listed in Table XVI for the two cases.  We see that there are enough candidates in both cases for three quark lepton families.   If we drop the assumption that the preonic doublet, one member of which binds in the Higgs, contributes its other member as a preonic singlet, then in both cases some of the entries in Table XVI become 0, and three complete families are no longer obtained.

\begin{table} [h]
\caption{Column 5 gives the number of Case (1) candidates, and column 6  gives the number of Case (2) candidates, for quark and lepton composites in the $SU(3)$, $SU(2)_{EW}$ and
$SU(2)_{TC}$ representations listed in columns 1 through 3,  with the $\overline{Y}$ value listed in column 4.   The notations D or S following the entries indicate respectively family doublet, singlet.}
\centering
\begin{tabular}{  c c c c c c}
\hline\hline

~~~$SU(3)$~~~ & ~~~$SU(2)_{EW}$~~~ & ~~~$SU(2)_{TC}$~~~& ~~~$\overline{Y}$~~~& ~~~Case (1) ~~~& ~~~ Case (2)~~~\\

\hline
 3 & 2& 1& $\frac{1}{6}$&27D,\,24S  &23D,\,24S \\
$\overline{3}$  &1 &2 &$-\frac{1}{6}$ &1D,\,1S  &1D,\,3S \\
 1 & 2& 1&$-\frac{1}{2}$ &2D,\,3S &1D,\,3S \\
 1 &1 & 2&$\frac{1}{2}$  &2D,\,1S  &1D,\,2S  \\

\hline\hline

\end{tabular}
\label{twelve}
\end{table}

\section{Anomaly matching conditions}

Since we are postulating that the massless preons bind to form massless composite quarks and leptons, we must address the 't Hooft anomaly
matching conditions \cite{thooft} that impose a consistency condition on the formation of such bound states.  Two different formulations of the
't Hooft conditions have been given in the literature, both starting from the assumption that there are one or more exact global chiral
symmetries, whose  conserved currents  can be coupled at the three vertices of an anomalous triangle graph.

't Hooft's argument
proposes gauging the global symmetries, which is possible once ``spectator'' fermions are added to the theory to cancel the global
triangle anomalies.  The low energy theory, in which some of the preons have been confined to bound states, must still be anomaly-canceling.
But since the spectator fermions contribute the same anomalies at high and low energies, the  anomalies computed at high energy  from preons
circulating in  each triangle graph must be equal to the  anomalies computed at low energy from composites circulating in the same triangle graph.
If anomalies do not match, then chiral symmetry must be spontaneously broken, with appearance of a corresponding Goldstone boson.

Frishman, Schwimmer, Banks, and Yankielowicz \cite{frish} , and Coleman and Grossman \cite{cole}, have given an alternative argument, which avoids the introduction of spectator fermions.  They proceed instead from the observation of Dolgov and Zakharov \cite{dol} that in a massless
chiral theory, the absorptive part of the anomalous triangle contains a $\delta(q^2)$ term.  In order for this term to match from the high
energy theory to the low energy theory, either the preonic and composite anomalies must match, or there must be a Goldstone boson arising
from chiral symmetry breaking, which also contributes a delta function to the absorptive part.

An essential feature of both arguments is that the global symmetry used to compute the anomaly must be exact.  Only an exactly conserved
current can be gauged, as in the 't Hooft argument, and exact conservation is needed to get the $\delta(q^2)$ in the absorptive part. Also,
the global symmetry must still involve only massless preons after the BEH mechanism gives some preons masses, since the
Dolgov--Zakharov calculation of a delta function in the absorptive part requires a limit of massless fermions.

Turning to the analysis of global symmetries in Sec. 3, we see that after the BEH mechanism gives masses to the Rarita-Schwinger fermions
in the 8 representation, only the charge $Q_2$ and its associated current obey the twin requirements of exact conservation and couplings
solely to massless fermions.  Note that after breaking of $SU(3) \times Sp(4)$ to $SU(3) \times SU(2) \times SU(2)$ there are many  approximate global symmetries
when couplings to the gauge fields that acquire masses through the BEH mechanism are neglected, but these do not correspond to
exactly conserved currents, so do not give rise to anomaly matching constraints.

To summarize, the only anomaly matching condition to be tested is the one coming from the $Q_2-Q_2-Q_2$ triangle graph. From Eq. \eqref{charges},
we have
\begin{equation}\label{anomaly}
8  \times{\rm Anomaly} (Q_2^3)=8  {\rm Anomaly}(N_{56}^3)-125  {\rm Anomaly}(N_{\overline{28}\perp}^3)~~~.
\end{equation}
We then have an exercise in counting.  For the preonic anomaly, we have
\begin{align}\label{preonic}
 {\rm Preonic~Anomaly}(N_{56}^3)=&56, ~~~{\rm Preonic~ Anomaly}(N_{\overline{28}\perp}^3)=28, ~~~\cr
  8\times {\rm Preonic~Anomaly}(Q_2^3)=&-3052~~~.\cr
\end{align}
For the composite anomaly, we have
\begin{align}\label{composite}
{\rm Composite~Anomaly}(N_{56}^3)=&\sum_{{\rm composite}~k} m_c(k)~\sum_{j\in 56}  N(k,j)^3~~~,\cr
{\rm Composite~Anomaly}(N_{\overline{28}\perp}^3)=& \sum_{{\rm composite}~k} m_c(k)~\sum_{j\in \overline{28}\perp} N(k,j)^3~~~,\cr
8\times {\rm Composite~Anomaly}(Q_2^3) =& 8{\rm Composite~Anomaly}(N_{56}^3)-125{\rm Composite~ Anomaly}(N_{\overline{28}\perp}^3)~~~.\cr
\end{align}
In Eq. \eqref{composite}, $m_c(k)=N_{SU(3)}(k) N_{SU(2)_{EW}}(k)N_{SU(2)_{TC}}(k)$ is the multiplicity of composite $k$, with the
$N$s the representation multiplicities in the first three columns of Tables XIII and XIV  (so $m_c(k)$ is 6 for quarks and 2 for leptons), and $N(k,j)$ is the number of times the preon $j$
appears in the composite $k$.  The computation time is greatly reduced by noting that the inner sums over $j$ for fixed composite label $k$
take a number of values versus $k$ far less than the number of candidates in Tables XIII and XIV, so it is necessary only to search for a match over the tables of these inner sum values, rather than over the much larger number of composite quark and lepton three family candidates.   The result after a search of close to $10^9$ possibilities, which took less than 10 seconds of computer time, is
that there is no three family match; the closest value of the composite anomaly found was  $8\times {\rm Composite~Anomaly}(Q_2^3)=-3060$.

This result means that the model can give rise to three composite families of quarks and leptons only if it saturates the anomaly matching
condition by also generating a Goldstone boson $B$ contribution, indicating that at some stage
of symmetry breaking, the chiral symmetry generated by $Q_2$ is spontaneously broken. Since $B$ contributes to the triangle through
a matrix element $\langle 0|Q_2|B\rangle ...$, with ... indicating the propagator and coupling of $B$ to the rest of the triangle, the quantum numbers of $B$ must match those of $Q_2$.  Since $Q_2$ is a gauge singlet with electric charge $Q=0$, the boson $B$ must be a singlet with
$Q=0$.   Also, since $B$ must be a scalar composite formed from left-handed chiral fields (the simplest is of the form $\Psi_{L1}^Ti\gamma^0\Psi_{L2}$), it must be chiral, that is, a scalar-pseudoscalar  mixture.  When radiative corrections
give  masses to the composite quarks and leptons, they will also be expected to give a mass to the Goldstone boson $B$, which then becomes
a pseudo-Goldstone boson, with possibly a very light mass.\footnote{The boson $B$ could be the axion \cite{weinberghomo} associated with the Peccei-Quinn
mechanism for solving the strong CP problem if the condensate that breaks the chiral symmetry generated by $Q_2$ also breaks chiral symmetry
generated by the global $U(1)$ that couples to the $SU(8)$ instanton.}  Interestingly, very light mass gauge singlet charge zero scalars are currently under active consideration \cite{hui} as dark matter candidates.

\section{Discussion, next steps, and experimental consequences}

In the proceeding paper (I) and this one we have started the analysis of the symmetry breaking chain in a new type of grand unified theory  \cite{adler1},\cite{adler3}.
The underlying motivation for a new attempt at grand unification is the fact that after over forty years of effort, an accepted  unification model that agrees with all experimental constraints has not been achieved.  This suggests that some essential ingredient in building a successful theory
has been overlooked, and that the rules for constructing unification models should be broadened.  The $SU(8)$ theory suggested in \cite{adler1}
incorporates a number of novel features.  It substitutes a principle of balance between boson and fermion degrees of freedom for a
requirement of full supersymmetry, and it allows spin-$\frac{3}{2}$ gauged Rarita-Schwinger fermions to play a role in anomaly cancellation,
instead of insisting that gauge anomalies cancel among the spin-$\frac{1}{2}$ fermions. The only scalar field in the model, a complex third
rank antisymmetric tensor, is forbidden by group theoretic considerations from having Yukawa couplings to the spin-$\frac{1}{2}$ fermions
of the model, a prerequisite \cite{weinberg3} for the model to give rise to a calculable low energy effective theory.

There are a number of  indications that the model of \cite{adler1} may be the correct one to explain observed standard model physics.
(1) As discussed in this paper, it contains a natural mechanism to remove the spin-$\frac{3}{2}$ particles from the low energy spectrum.
(2) The model does not employ a Higgs potential with parameters that have to be chosen to give the observed symmetry breaking pattern. Symmetry
breaking is initiated by the Coleman-Weinberg mechanism for the scalar field, and the pattern of symmetry breaking is determined
by the kinematics of the gauge boson mass matrix.  (3) As noted in (I), symmetry breaking by a third rank antisymmetric tensor gives a
natural mechanism for emergence of $SU(3)$ as the color group, since a rank three antisymmetric tensor provides an $SU(3)$ invariant.
(4)  As observed in this paper,  the co-running of the $SU(3)$ and $Sp(4)$ running couplings that emerges naturally after $SU(8)$
symmetry breaking is a prerequisite for realistic dynamical symmetry breaking at lower energies. (5) As discussed in detail in this paper, the model
has a possible symmetry breaking chain leading to the particles and forces  of the standard model.

Many open questions remain.  In this paper we have focused on qualitative kinematic aspects of symmetry breaking that can addressed by group theory, one-loop running coupling analysis, enumeration of composite candidates, and anomaly counting.  To answer more detailed quantitative questions such as  the  mass scales associated with the
various stages of symmetry  breaking, the binding of composites, and the generation of fermion masses, will require computation of higher order radiative corrections, a task for the future.
The most pressing issues are (1) seeing whether radiative corrections to the Coleman-Weinberg potential for the scalar field lead
to a transition at a scale $M_{Sp(4)}$ to a phase with asymmetric breaking of $Sp(4)$ into $SU(2)_{TC} \times SU(2)_{EW}$, and (2) seeing whether
radiative corrections to the vector meson mass matrix place the theory  within a ``conformal window'', so that the model evolves in the
infrared to a fixed point at a scale $M_U$, below which there is an extra $U(1)$ symmetry with generator $G$.  Both of these requirements are needed
for the model to give rise to the standard model as its low energy effective field theory.

However, based on the analysis of this paper, one can already state some experimental consequences should the $SU(8)$ model prove
theoretically viable:

\begin{enumerate}

\item {\it Composite structure} The model implies that quarks and leptons are composites formed from more fundamental preons.  However, unlike earlier
preon models, such as the ones reviewed by Fritzsch \cite{fritzsch} and Wesenberg \cite{wesen}, where the aim is to have a small number of preons, our
model starts from a grand unified framework with large 56 and $\overline{28}$ multiplets of preons. Because our model is based on grand unification
of the gauge forces,  the weak bosons are elementary gauge bosons rather than composites as in \cite{fritzsch} and \cite{wesen}, with only the Higgs
and possibly a very light scalar boson appearing as composites.

Experimentally, composite quarks and leptons are a viable possibility at an energy scale well above that attainable at the LHC.
A generic feature of compositeness
is that the distinction between the strong and electroweak interactions gets blurred as the compositeness scale is approached.  In this regard,
it is interesting that the Auger experiment \cite{auger} has reported an excess of muons produced at center of mass energies of order 100-170 TeV, which could be indicative of changes in the particle forces at energies well above the LHC range.

\item {\it Proton stability}  The composite quarks and leptons in the model have different
internal structure:  leptons are three preon composites, and quarks are five preon composites.  Thus  proton decay into a single lepton
plus mesons involves a large change in preon number.
We have seen that in low energy processes,   $Q_2=N_{56} - (5/2) N_{\overline{28}\perp}$  and  $\tilde{Q}_1=N_{\overline{28}\Delta \lambda} -
(3/5) N_{56}$ are conserved,
which implies that the changes  $\delta(N_{56})$, $\delta(N_{\overline{28}\perp})$, and
$\delta(N_{\overline{28}\Delta \lambda})$ in a decay are related by
\begin{align}\label{pdecay}
\delta(N_{56})=&\frac{5}{3}\delta(N_{\overline{28}\Delta \lambda})~~~,\cr
\delta(N_{\overline{28}\perp})=&\frac{2}{3}\delta(N_{\overline{28}\Delta \lambda})~~~.\cr
\end{align}
In general these forbid the preon number changes required in proton decay, so the proton in our model is stable against decay into a
single lepton plus mesons.

\item {\it Higgs field}  The Higgs field in the model can be an $SU(2)_{TC}$ singlet, which is the favored possibility for binding, but
not an $SU(2)_{TC} \times SU(2)_{EW}$ bi-doublet.

\item{\it Dark matter}  The model requires that the three standard model families of fermions be accompanied by one or more scalar
pseudo-Goldstone bosons to satisfy the anomaly matching condition.  This favors a very light boson explanation of cosmological dark matter, such
as that discussed in \cite{hui}.

\end{enumerate}

\section{Acknowledgements}
I wish to thank Peter Goddard and Paul Langacker for helpful conversations about index normalization and the conventions for
Casimir invariants.  Work on this paper was supported in part by the National Science Foundation under Grant
No. PHYS-1066293 and the hospitality of the Aspen Center for Physics.

\appendix
\section{$U(1)$ matching relation}

To derive the matching relation of Eq. \eqref{matching}, we start from the coupling
term
\begin{equation}\label{coupling1}
g_{TC} A_{TC} T_{3TC}+ g_U A_U G~~~,
\end{equation}
with $A_{TC}$ and $A_{U}$ the time components of the respective gauge fields,
both of which have standard kinetic energy normalization.  Since $Y=T_{3TC}+G$,
we look for a rotated set of gauge fields for which Eq. \eqref{coupling1} takes
the form
\begin{equation}\label{coupling2}
g^{\prime} (A_{TC} \cos \theta + A_U \sin \theta) (T_{3TC}+G)
+ (-A_{TC}\sin\theta + A_U \cos \theta) (u T_{3TC} + v G)~~~,
\end{equation}
with $g^{\prime}$, $\theta$, $u$, and $v$ determined by matching Eq. \eqref{coupling2}
to Eq. \eqref{coupling1}.  The matching conidtions are
\begin{align}
g^{\prime} \cos \theta-v\sin\theta=&0~~~,\cr
g^{\prime} \sin \theta+u \cos \theta=&0~~~,\cr
g^{\prime} \cos \theta-u \sin \theta=&g_{TC}~~~,\cr
g^{\prime} \sin \theta+ v \cos \theta=&g_U~~~.\cr
\end{align}
Solving these gives
\begin{align}\label{solution1}
g^{\prime}=&g_{TC} \cos \theta=g_U \sin \theta~~~,\cr
u=&-g_{TC} \sin \theta ~,~~v=g_U \cos \theta~~~,\cr
\tan \theta= &g_{TC}/g_U ~~~,\cr
\end{align}
from which we get
\begin{equation}\label{solution2}
\frac{1}{g^{\prime\,2}}= \frac{1}{g_{TC}^2}(1+\tan^2 \theta)
=\frac{1}{g_{TC}^2}+\frac{1}{g_U^2}~~~.
\end{equation}


\begin{thebibliography}{99}
\bibitem{adler1}  S. L. Adler, Int. J. Mod. Phys. A {\bf 29}, 1450130 (2014).
\bibitem{adler3}  S. L. Adler, J. Phys. A: Math. Theor. {\bf 49}, 315401 (2016).  This paper is referred
to as (I) after first cited in the text.
\bibitem{adler2} S. L. Adler, Phys. Rev. D {\bf 92}, 085022 and 085023 (2015).
\bibitem{shifman} M. Shifman, {\it Instantons in Gauge Theories}, World Scientific, Singpore (1994), p. 214.
\bibitem{okubo} S. Okubo, J. Math. Phys. {\bf 26}, 2127 (1985), Eqs. (1.16)--(1.18c).
\bibitem{mckay}  W. G. McKay and J. Patera, {\it Tables of Dimensions, Indices,
and Branching Rules for Representations of Simple Lie Algebras}, Marcel Dekker, New York
and Basel (1981).  The index in their tables is given in the column $I(2)/n$ with $n$ the group rank.
\bibitem{goddard} P. Goddard and D. Olive, Int. J. Mod. Phys. A {\bf 1}, 303 (1986), p. 363.
\bibitem{slansky}  R. Slansky, Phys. Reports  {\bf 79}, 1 (1981).
\bibitem{zee} For reprints of the seminal $SU(5)$, $SO(10)$, and Pati-Salam model papers, see A. Zee, {\it Unity of Forces in
the Universe}, Vol. I, World Scientific, Singapore (1982). For a good exposition, see Chapters 5-7 of \cite{mohapatra}.
\bibitem{barr} S. M. Barr, Phys. Lett. B {\bf 112}, 219 (1982).  For some subsequent references,
see \cite{adler1}.
\bibitem{mohapatra} R. N. Mohapatra, {\it Unification and Supersymmetry}, Chapter 6, Springer-Verlag, New York (1992).
\bibitem{preskill} J. P. Preskill, Phys. Rev. Lett. {\bf 43}, 1365 (1979).
\bibitem{weinberghomo} S. Weinberg, {\it The Quantum Theory of Fields, Vol. II Modern Applications}, Cambridge University Press,
Cambridge (1996).  For homotopy groups, see  p. 442 and Appendix B,  for axions and the Peccei-Quinn mechanism, see  pp. 458-461,
and for running of the standard model couplings see pp. 329-330.
\bibitem{weinberg1} S. Weinberg,   Phys. Rev. D{\bf 13}, 974 (1976).  See p. 92 for a discussion of the need for similar running rates of the
strong coupling and the coupling in the sector giving rise to dynamical symmetry breaking.
\bibitem{susskind} L. Susskind, Phys. Rev. D{\bf 20}, 2619 (1979).
\bibitem{weinberg2}S. Weinberg,  Phys. Rev. D{\bf 19}, 127 (1979).
\bibitem{lane}  K. Lane, ``An  introduction to technicolor'', arXiv:hep-ph/9401324.
\bibitem{piai} M. Piai, Adv. High Energy Phys. 2010:464302 (2010).
\bibitem{schmaltz} M. Schmaltz and D. Tucker-Smith, Ann. Rev. Nucl. Part. Sci. {\bf 55}, 229 (2005).
\bibitem{perelstein} M. Perelstein, Prog. Part. Nucl. Phys. {\bf 58}, 247 (2007).
\bibitem{raby}  S. Raby, S. Dimopoulos, and L. Susskind, Nucl. Phys. B {\bf169}, 373 (1980).
\bibitem{peskin}M. Peskin, ``Compositeness of Quarks and Leptons'',~  http://lss.fnal.gov/conf/C810824/p880.pdf ~(1981).
\bibitem{fritzsch}  H. Fritzsch, Mod. Phys. Lett. A {\bf 26}, 2305 (2011).
\bibitem{wesen} G. H. Wesenberg, ``Preon Models in Particle Physics'', NTNU Masters Thesis (2014).
\bibitem{thooft}  G. 't Hooft, in {\it Recent developments in gauge theories} (1979 Cargeses lectures), ed. G. 't Hooft et al., Plenum, New
York (1980).
\bibitem{frish} Y. Frishman, A. Schwimmer, T. Banks, and S. Yankielowicz, Nucl. Phys. B{\bf 177}, 157 (1981).
\bibitem{cole} S. Coleman and B. Grossman, Nucl. Phys. B {\bf 203}, 205 (1982).
\bibitem{dol} A. D. Dolgov and V. I. Zakharov, Nucl. Phys. B{\bf 27},525 (1971).
\bibitem{hui} L. Hui, J. P. Ostriker, S. Tremaine, and E. Witten, Phys. Rev. D {\bf 95}, 043541 (2017).  
\bibitem{weinberg3} S. Weinberg, Phys. Rev. Lett. {\bf 29}, 388 (1972).
\bibitem{auger} A. Aab et. al, Phys. Rev. Lett. {\bf 117}, 192001 (2016).
\end{thebibliography}
\end{document}